\documentclass[conference]{IEEEtran}

\usepackage{amsmath,amsfonts,amssymb}
\usepackage{graphicx}
\usepackage{bm}
\usepackage{cite}
\usepackage{url}
\usepackage{color}
\usepackage{svg}

\title{
Safe Sliding Mode Control for Marine Vessels Using
High-Order Control Barrier Functions and Fast Projection
}

\author{
\IEEEauthorblockN{Spyridon Syntakas, Kostas Vlachos}
}

\begin{document}
\maketitle

\begin{abstract}
This paper presents a novel safe control framework that integrates 
Sliding Mode Control (SMC), High-Order Control Barrier Functions (HOCBFs) with state-dependent adaptiveness
and a lightweight projection for collision-free
navigation of an over-actuated 3-DOF marine surface vessel subjected
to strong environmental disturbances (wind, waves, and current).
SMC provides robustness to matched disturbances common in marine operations,
while HOCBFs enforce forward invariance of obstacle-avoidance constraints.
A fast half-space projection method adjusts the SMC control only when needed,
preserving robustness and minimizing chattering.
The approach is evaluated on a nonlinear marine platform model that includes
added mass, hydrodynamic damping, and full thruster allocation.
Simulation results show robust navigation, guaranteed obstacle
avoidance, and computational efficiency suitable for real-time embedded use. For small marine robots and surface vessels with limited onboard computational resources—where execution speed and computational efficiency are critical—the SMC–HOCBF framework constitutes a strong candidate for safety-critical control.
\end{abstract}

\section{Introduction}

Autonomous marine vessels must operate safely under strong disturbances
and nonlinear hydrodynamics while navigating around obstacles.
Sliding Mode Control (SMC) \cite{survey_smc} is widely known for its robustness to
matched disturbances, making it attractive for surface vessels exposed to
wind, waves, and current.
However, SMC provides no inherent safety guarantees,
creating risk in  environments with obstacles.

Control Barrier Functions (CBFs) \cite{ames2019controlbarrierfunctionstheory} have recently emerged as a powerful tool
for enforcing safety constraints using forward set invariance.
While CBFs are popular in robotics and autonomous driving,
their integration with SMC is not common, especially for fully nonlinear,
hydrodynamically consistent marine vessels, while some work has been conducted for systems with a more academic focus \cite{LAGHROUCHE2021109355}. For instance, a theoretical treatment of SMC with high-order barrier functions is done in \cite{Chinelato2022}, \cite{chinelato2020slidingmodecontrolbarrier}. While \cite{Chinelato2022} focuses is on the theoretical development of SMC–CBF methods, the current work applies an SMC–HOCBF controller to a hydrodynamically complex marine platform and benchmarks it against a tube-based NMPC under identical disturbance scenarios.
Furthermore, solving the standard CBF quadratic program (QP)
may be computationally heavy for embedded marine systems.

This work introduces a control architecture that:
\begin{itemize}
\item uses a \textbf{sliding-mode control law} for robust tracking 
\item employs a \textbf{high-order control barrier function} (HOCBF) for obstacle avoidance with \textbf{state-dependent adaptiveness}
\item applies a \textbf{fast projection operator} that replaces the full QP,
\item and is implemented on a fully nonlinear \textbf{3-DOF marine dynamics model}.
\end{itemize}

The resulting controller guarantees safety in the sense of forward
invariance of a safe set, while preserving robust tracking properties
typical of SMC. For small marine robots and surface vessels with constrained onboard processors—where real-time performance and computational efficiency are paramount—the proposed SMC–CBF framework emerges as a compelling solution for safety-critical control. Its purely analytical structure, combined with lightweight safety enforcement, enables reliable disturbance rejection and constraint satisfaction without the overhead associated with nonlinear optimization–based approaches.

\section{Related Work}

\subsection{Control Barrier Functions and Safety-Critical Control}

Control Barrier Functions (CBFs) \cite{ames2019controlbarrierfunctionstheory} have become a standard tool for
safety-critical control in robotics and autonomous systems, providing
conditions for forward invariance of a safe set via inequality
constraints on the control input.
CBFs are typically combined with a nominal controller through a Quadratic Program (QP) that minimally modifies the nominal input to ensure satisfaction of safety constraints.
This CBF--QP paradigm has been widely adopted in mobile robotics,
autonomous driving, and legged locomotion.

High-Order CBFs (HOCBFs) \cite{high-ord} generalize this framework to outputs with
relative degree greater than one, allowing position-level safety
constraints in systems whose control inputs appear only after multiple
derivatives of the state.

\subsection{SMC and Robust Control}

Sliding Mode Control (SMC) is a well-established robust control method
for nonlinear systems with matched disturbances \cite{survey_smc}.
In the marine domain, SMC and related robust controllers have been
studied for station keeping, dynamic positioning, and path following
under environmental disturbances such as wind, waves, and current
\cite{shippositioning}. These works exploit the robustness
properties of SMC with respect to unmodeled dynamics and bounded
external forces, but do not typically incorporate explicit state
constraints or formal safety guarantees such as collision avoidance with
obstacles.

\subsection{CBF-Based MPC and Safe Navigation}

Model Predictive Control (MPC) has been widely used for constrained
trajectory planning, and its combination with CBFs has recently been
explored to enforce safety in a predictive framework
\cite{zeng2021mpccbf}, \cite{TUBEMPC_SYNTAKAS}.
In these approaches, safety constraints derived from CBFs or geometric
distance functions are enforced over a finite horizon, often at the cost
of solving relatively large QPs or nonlinear programs at each sampling
time.
Although effective, such methods may be computationally demanding for
embedded marine platforms and may not directly exploit the disturbance
rejection capabilities of SMC.

\subsection{Sliding Mode Control with CBFs}

A few recent works have begun to explore the combination of SMC with
CBFs for safety-critical control, e.g., \cite{smccbf1}, \cite{smccbf2}, \cite{LAGHROUCHE2021109355}.
In general, methods typically formulate a CLF--CBF--QP where an SMC-like term
appears in the cost or dynamics, and a full QP is solved online to
enforce CBF constraints.
However, such works are primarily developed for quadrotor or low-order
systems with simplified dynamics and do not address fully nonlinear
marine vessel models with hydrodynamic coupling, realistic environmental
disturbances, and thruster allocation.

\subsection{Contributions Relative to Prior Work}

Compared to the above literature, the main contributions of this work
are:
\begin{itemize}
\item We develop a \emph{safe sliding-mode control} framework for an over-actuated 3-DOF marine vessel subject to realistic wind, wave, and
current disturbances, using a high-fidelity nonlinear model rather than
simplified kinematics.
\item We integrate a high-order CBF for circular obstacle avoidance
directly into the body-force control space, capturing the relative
degree two structure of position-level safety constraints in the marine
dynamics.
\item We replace the standard CBF--QP with a \emph{fast projection-based
safety filter} that iteratively projects the nominal SMC control onto
the intersection of half-space CBF constraints and actuator bounds,
reducing computational complexity and making the method more suitable
for real-time embedded marine control.
\item We provide a Lyapunov-based analysis establishing practical
stability of the sliding surface and forward invariance of the safe set
under environmental disturbances, explicitly characterizing the
robustness and safety properties of the combined SMC--HOCBF--projection
architecture.
\end{itemize}

To the best of the authors' knowledge, this is the first work to
combine sliding-mode tracking, high-order CBF safety constraints, and a
lightweight projection operator in the context of nonlinear 3-DOF marine
vessel dynamics with wind, wave, and current disturbances.

\section{Marine Vessel Dynamics}

The vessel used in the simulations of this work is a overactuted marine platform that serves as a representative
case study for the proposed control methodology.  The system was originally developed
as an auxiliary autonomous surface vehicle intended to support a deep-sea neutrino
telescope; a complete description of its design and hydrodynamic characteristics is
provided in \cite{VLACHOS201510}. The vessel has a mass of 
\(425\times10^{3}\,\mathrm{kg}\) and features an isosceles–triangle geometry equipped 
with three fully actuated azimuth thrusters located at its vertices. Each thruster is 
housed within a hollow double–cylinder structure and is capable of delivering 
vectorized forces up to \(20\,\mathrm{kN}\). The units operate in a fully submerged 
configuration, rotating parallel to the free surface through electro–hydraulic 
actuation.

Several control strategies have been previously investigated for this platform, from standard \cite{VLACHOS201510} to learning based methodologies \cite{Syntakas_2023}. Of particular relevance to the  present work, a robust tube–based NMPC scheme incorporating Control Barrier Functions 
for safe obstacle avoidance was developed in \cite{TUBEMPC_SYNTAKAS}. This controller is adopted here as a benchmark for comparison with the proposed method.

Finally, it is noted that the actuator dynamics of the azimuth thrusters evolve on a 
significantly faster timescale than the rigid–body vessel dynamics. This separation 
of timescales ensures that the thruster behavior does not limit the performance of 
the proposed controller nor the thrust–allocation scheme employed in this study.

We adopt the standard 3-DOF nonlinear marine model \cite{bookFOSSEN}, \cite{Fossen1994GuidanceAC_disturbances}.
Let
\[
\bm{\eta} = [x, y, \psi]^T, \quad 
\bm{\nu} = [u, v, r]^T
\]
represent the pose and body-fixed velocities.
The kinematics are
\begin{equation}
\dot{\bm{\eta}} = \bm{R}(\psi)\bm{\nu},
\end{equation}
with the rotation matrix
\[
\bm{R}(\psi) =
\begin{bmatrix}
\cos\psi & -\sin\psi & 0 \\
\sin\psi & \cos\psi & 0 \\
0 & 0 & 1
\end{bmatrix}.
\]

The dynamics are:
\begin{equation}
\bm{M}\dot{\bm{\nu}}
+ \bm{C}(\bm{\nu})\bm{\nu}
+ \bm{D}(\bm{\nu})\bm{\nu}
= \bm{\tau} + \bm{d},
\label{eq:dynamics}
\end{equation}
where:
\begin{itemize}
\item $\bm{M}$ is the positive definite mass + added-mass matrix,
\item $\bm{C}(\bm{\nu})$ is the Coriolis-centripetal matrix,
\item $\bm{D}(\bm{\nu})$ is hydrodynamic damping,
\item $\bm{d}(t)$ represents disturbance forces due to wind, waves, and current,
\item $\bm{\tau} = [\tau_x, \tau_y, \tau_n]^T$ are the control inputs.
\end{itemize}

We assume:

\medskip
\noindent
\textbf{Assumption 1.}
The matrices $\bm{M}$ and $\bm{D}(\bm{\nu})$ are bounded, 
$\bm{M}$ is positive definite,
and $\bm{C}(\bm{\nu})$ satisfies the skew-symmetry property
$\bm{\nu}^T(\bm{C}(\bm{\nu}) + \bm{C}^T(\bm{\nu}))\bm{\nu} = 0.$

\medskip
\noindent
\textbf{Assumption 2.}
The disturbance vector $\bm{d}(t)$ is bounded:
$\|\bm{d}(t)\|\le d_{\max}$ for all $t\ge 0$,
and enters the system in the same channels as the control input
(i.e., it is \emph{matched}).

\section{Environmental Disturbances}

\begin{figure}[!htbp]
    \centering
    \centering
    \includegraphics[width=0.95\linewidth]{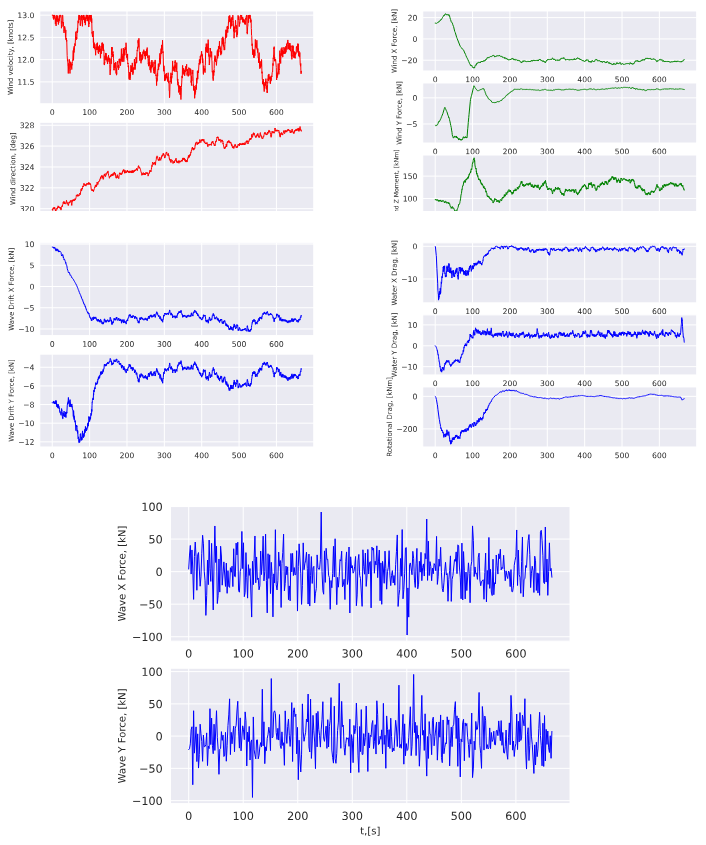}
    \caption{Environmental disturbances used in simulations.}
    \label{fig:disturbances}
\end{figure}

In this work, the vessel operates under realistic bounded environmental
disturbances generated by separate physical models for wind, waves, and
sea current, that are based on measurements from the Mediterranean Sea. A representative depiction is seen in Fig. \ref{fig:disturbances}. The total disturbance vector in \eqref{eq:dynamics} is
decomposed as
\begin{equation}
\bm{d}(t) = \bm{d}_{\mathrm{wind}}(t)
+ \bm{d}_{\mathrm{wave}}(t)
+ \bm{d}_{\mathrm{curr}}(t),
\end{equation}
where:
\begin{itemize}
\item $\bm{d}_{\mathrm{wind}}(t)$ models aerodynamic forces and moment
induced by wind acting on the exposed structure,
\item $\bm{d}_{\mathrm{wave}}(t)$ includes a low-frequency drift component
and a zero-mean stochastic component due to irregular waves,
\item $\bm{d}_{\mathrm{curr}}(t)$ represents hydrodynamic loads from sea current.
\end{itemize}

In the implementation, these components are generated using external
disturbance models, and are applied in the same channels as the control
inputs, i.e., they are matched disturbances. Their combined effect is
bounded, consistent with Assumption~2. The proposed Sliding-Mode
Control--CBF--projection architecture is therefore explicitly designed and
evaluated under non-zero, time-varying environmental disturbance, and
the robustness results in the stability analysis refer to this
disturbed operating regime rather than idealized disturbance-free
conditions.

\section{Sliding-Mode Control Design}

Let $\bm{\eta}_d$ be a desired goal, and define:
\[
\bm{e}_p = \bm{\eta} - \bm{\eta}_d, 
\qquad
\bm{e}_v = \bm{\nu} - \bm{\nu}_d,
\]
where $\bm{\nu}_d$ is a desired velocity.

\subsection{Sliding Surface}

We define the sliding surface:
\begin{equation}
\bm{s} = \bm{e}_v + \bm{\Lambda}\bm{e}_p,
\label{eq:sdef}
\end{equation}
where $\bm{\Lambda} \in \mathbb{R}^{3\times 3}$ is diagonal and positive definite.

Differentiating \eqref{eq:sdef}:
\begin{align}
\dot{\bm{s}} 
&= \dot{\bm{e}}_v + \bm{\Lambda}\dot{\bm{e}}_p \nonumber \\
&= \dot{\bm{\nu}} - \dot{\bm{\nu}}_d + \bm{\Lambda}(\bm{R}\bm{\nu} - \dot{\bm{\eta}}_d).
\end{align}

Using \eqref{eq:dynamics}, we have:
\begin{align}
\dot{\bm{\nu}}
&= \bm{M}^{-1}(\bm{\tau} + \bm{d} - \bm{C}\bm{\nu} - \bm{D}\bm{\nu}).
\end{align}

Hence
\begin{align}
\dot{\bm{s}} 
&= \bm{M}^{-1}(\bm{\tau} + \bm{d} - \bm{C}\bm{\nu} - \bm{D}\bm{\nu})
- \dot{\bm{\nu}}_d \nonumber\\
&\quad + \bm{\Lambda}\bm{R}\bm{\nu} - \bm{\Lambda}\dot{\bm{\eta}}_d.
\label{eq:s_dot_expanded}
\end{align}

\subsection{SMC Law}

We choose a boundary-layer sliding dynamics:
\begin{equation}
\dot{\bm{s}} = -\bm{K}_s\, \mathrm{sat}\!\left(\frac{\bm{s}}{\phi}\right),
\label{eq:s_desired}
\end{equation}
where $\bm{K}_s$ is diagonal positive definite,
$\phi>0$ is the boundary layer thickness, and
$\mathrm{sat}(\cdot)$ is applied elementwise.

Equating \eqref{eq:s_dot_expanded} and \eqref{eq:s_desired} and solving for $\bm{\tau}$:
\begin{align}
\bm{\tau}
&=
\bm{C}\bm{\nu}
+ \bm{D}\bm{\nu}
- \bm{d}
+ \bm{M}
\Big(
 -\bm{K}_s\mathrm{sat}\!\left(\frac{\bm{s}}{\phi}\right)
+ \dot{\bm{\nu}}_d \nonumber\\
&\quad\quad
- \bm{\Lambda}\bm{R}\bm{\nu}
+ \bm{\Lambda}\dot{\bm{\eta}}_d
\Big).
\label{eq:tau_smc_full}
\end{align}

In implementation, the disturbance $\bm{d}$ is not exactly known.
We write the implementable control as:
\begin{equation}
\bm{\tau}_{\text{SMC}} =
\bm{C}\bm{\nu}
+ \bm{D}\bm{\nu}
+ \bm{M}
\Big(
 -\bm{K}_s\mathrm{sat}\!\left(\frac{\bm{s}}{\phi}\right)
+ \dot{\bm{\nu}}_d
- \bm{\Lambda}\bm{R}\bm{\nu}
+ \bm{\Lambda}\dot{\bm{\eta}}_d
\Big)
\label{eq:tau_smc_impl}
\end{equation}
and treat $-\bm{d}$ as a bounded additive perturbation.

\section{High-Order Control Barrier Function}

We consider a circular obstacle centered at $(x_o,y_o)$ with 
safety radius $R>0$.
Define:
\begin{equation}
h(\bm{\eta}) = (x-x_o)^2 + (y-y_o)^2 - R^2.
\end{equation}
The safe set is:
\[
\mathcal{C} = \{\bm{\eta} \in \mathbb{R}^3 \mid h(\bm{\eta}) \ge 0\}.
\]

\subsection{Relative Degree and HOCBF}

The function $h$ depends only on $\bm{\eta}$.
Using $\dot{\bm{\eta}} = \bm{R}\bm{\nu}$,
\begin{equation}
\dot{h} = \frac{\partial h}{\partial \bm{\eta}}\dot{\bm{\eta}}
= 2(x-x_o)\dot{x} + 2(y-y_o)\dot{y}.
\end{equation}
Since $\dot{x},\dot{y}$ depend on $\bm{\nu}$ but not directly on $\bm{\tau}$,
and $\dot{\bm{\nu}}$ depends on $\bm{\tau}$,
the input $\bm{\tau}$ appears first in $\ddot{h}$.
Hence $h$ has \textit{relative degree two} with respect to the control.

A High-Order Control Barrier Function (HOCBF) \cite{high-ord}
enforces safety via:
\begin{equation}
\ddot{h} + 2\alpha \dot{h} + \alpha^2 h \ge 0,
\label{eq:hocbf_cond}
\end{equation}
for some $\alpha>0$.

Using Lie derivatives, this can be written as:
\begin{equation}
L_f^2 h + L_g L_f h\, \bm{\tau}
+ 2\alpha L_f h
+ \alpha^2 h \ge 0.
\end{equation}
This yields a linear inequality in $\bm{\tau}$:
\begin{equation}
\bm{A}\bm{\tau} \ge \bm{b},
\label{eq:CBF_constraint}
\end{equation}
where $\bm{A}\in\mathbb{R}^{1\times 3}$, $\bm{b}\in\mathbb{R}$ depend on the current state.

\medskip
\noindent
\textbf{Theorem 1 (Forward Invariance of the Safe Set).}
\textit{
Suppose $h$ is twice continuously differentiable, has relative degree two,
and the HOCBF condition \eqref{eq:hocbf_cond} holds for all $t\ge 0$.
If $h(\bm{\eta}(0))\ge 0$, then $h(\bm{\eta}(t))\ge 0$ for all $t\ge 0$.
}

\medskip
\noindent
\textit{Proof:}
This is a standard consequence of HOCBF theory \cite{high-ord}:
the inequality \eqref{eq:hocbf_cond} enforces exponential lower bounds on $h$,
ensuring that solutions starting in $\mathcal{C}$ remain in $\mathcal{C}$.
\hfill$\blacksquare$

\section{Projection-Based Safety Filter}

We define a projection operator that maps an arbitrary input 
$\bm{\tau}$ to a nearby safe control, avoiding the need to solve a full 
quadratic program at each step. This keeps the computational load 
sufficiently low for systems equipped with limited onboard processing 
capabilities.

Starting from the nominal sliding-mode control 
$\bm{\tau}_{\text{SMC}}$ in \eqref{eq:tau_smc_impl}, we enforce the CBF 
constraint \eqref{eq:CBF_constraint} together with the actuator bounds
\[
    \bm{\tau}_{\min} \le \bm{\tau} \le \bm{\tau}_{\max}.
\]

The corresponding safety filter may be expressed as the projection
\begin{equation}
\bm{\tau}^\star
= \arg\min_{\bm{\tau}} \|\bm{\tau}-\bm{\tau}_{\text{SMC}}\|^2
\quad
\text{s.t. }\bm{A}\bm{\tau}\ge\bm{b},\;
\bm{\tau}_{\min}\le\bm{\tau}\le\bm{\tau}_{\max},
\label{eq:QP_filter}
\end{equation}
but rather than solving \eqref{eq:QP_filter} directly, we employ a fast 
sequential projection scheme.

We initialize
\[
    \bm{\tau}^{(0)} = 
    \Pi_{\mathcal{U}}(\bm{\tau}_{\text{SMC}}),
    \qquad 
    \mathcal{U} = 
    \{\bm{\tau}\mid 
      \bm{\tau}_{\min}\le\bm{\tau}\le\bm{\tau}_{\max}\},
\]
where $\Pi_{\mathcal{U}}$ denotes componentwise clipping.  
For each barrier constraint $\bm{a}_j^\top\bm{\tau} \ge b_j$, we compute 
the violation
\[
    v_j = b_j - \bm{a}_j^\top\bm{\tau}^{(k)}.
\]
Whenever $v_j>0$, a relaxed projection step is applied:
\begin{equation}
\bm{\tau}^{(k)}
\leftarrow
\bm{\tau}^{(k)} 
+ 
\gamma\,
\frac{v_j}{\|\bm{a}_j\|^2}\,
\bm{a}_j,
\qquad 
\gamma\in(0,1],
\label{eq:relaxed_projection}
\end{equation}
followed by actuator clipping 
$\bm{\tau}^{(k)}\!\leftarrow\!\Pi_{\mathcal{U}}(\bm{\tau}^{(k)})$.
A small number of sweeps over all rows of $\bm{A}$ yields a control input 
satisfying all CBF and actuator constraints to within a prescribed 
tolerance.

We denote the resulting safe input by
\begin{equation}
\bm{\tau}_{\text{safe}}
= 
\bm{\tau}^{(K)}.
\end{equation}
This projected input is the one applied to the system:
\begin{equation}
\bm{\tau} = \bm{\tau}_{\text{safe}}.
\end{equation}

\section{Stability and Safety Analysis}

We now give a more detailed analysis of tracking and safety under the
proposed controller.

\subsection{SMC Tracking without CBF Constraints}

We first consider the ideal case where the CBF constraint is inactive,
i.e., $\bm{\tau} = \bm{\tau}_{\text{SMC}}$ from \eqref{eq:tau_smc_impl}.

\medskip
\noindent
\textbf{Theorem 2 (SMC Tracking with Disturbance).}
\textit{
Assume that $\bm{\nu}_d$ and $\bm{\eta}_d$ are bounded and sufficiently
smooth, and that the disturbance $\bm{d}(t)$ is matched and bounded as
in Assumption~2, i.e.\ $\|\bm{d}(t)\|\le d_{\max}$ for all $t\ge 0$.
Let $\lambda_{\min}(\bm{M})$ denote the smallest eigenvalue of the
positive definite matrix $\bm{M}$, and let
\(
    d_{\mathrm{eq}} := d_{\max}/\lambda_{\min}(\bm{M}).
\)
If the SMC gain matrix $\bm{K}_s$ satisfies
\[
    \lambda_{\min}(\bm{K}_s) > d_{\mathrm{eq}},
\]
then the sliding surface $\bm{s}$ in \eqref{eq:sdef} is globally bounded
and \emph{ultimately bounded} in a neighborhood of the origin, i.e.,
there exists a radius $\delta>0$ such that
\[
    \limsup_{t\to\infty} \|\bm{s}(t)\| \le \delta,
\]
where $\delta$ can be made arbitrarily small by increasing
$\lambda_{\min}(\bm{K}_s)$ and decreasing the boundary-layer thickness
$\phi$, subject to actuator limits.
}

\medskip
\noindent
\textit{Proof:}
Define the Lyapunov function
\[
    V(\bm{s}) = \tfrac{1}{2}\,\bm{s}^\top \bm{s}.
\]
Using the dynamics of $\bm{s}$ and the ideal SMC law
\eqref{eq:tau_smc_full}, one obtains
\[
    \dot{\bm{s}}
    = -\bm{K}_s \,\mathrm{sat}\!\left(\frac{\bm{s}}{\phi}\right),
\]
and hence
\[
    \dot{V}
    = \bm{s}^\top \dot{\bm{s}}
    = -\bm{s}^\top \bm{K}_s \,
      \mathrm{sat}\!\left(\tfrac{\bm{s}}{\phi}\right)
    \le 0,
\]
showing Lyapunov stability in the absence of disturbances.

In practice, the disturbance $\bm{d}(t)$ is not exactly canceled and
enters the $\bm{s}$-dynamics as a matched perturbation. Using
\eqref{eq:dynamics} together with the implementable control
\eqref{eq:tau_smc_impl}, the sliding dynamics can be written as
\[
    \dot{\bm{s}}
    = -\bm{K}_s \,\mathrm{sat}\!\left(\tfrac{\bm{s}}{\phi}\right)
      + \bm{\Delta}(t),
\]
where $\bm{\Delta}(t) = \bm{M}^{-1}\bm{d}(t)$ and
\[
    \|\bm{\Delta}(t)\|
    \le \frac{d_{\max}}{\lambda_{\min}(\bm{M})}
    = d_{\mathrm{eq}}
    \quad \forall t\ge 0.
\]

Differentiating $V$ along trajectories gives
\[
    \dot{V}
    = \bm{s}^\top \dot{\bm{s}}
    = -\bm{s}^\top \bm{K}_s
      \,\mathrm{sat}\!\left(\tfrac{\bm{s}}{\phi}\right)
      + \bm{s}^\top \bm{\Delta}(t).
\]

For each component $s_i$, the saturation satisfies
\[
s_i\,\mathrm{sat}(s_i/\phi) \;\ge\; |s_i| - \phi,
\]
and since $\bm{K}_s$ is positive diagonal,
\[
\bm{s}^\top \bm{K}_s \,\mathrm{sat}\!\left(\tfrac{\bm{s}}{\phi}\right)
\;\ge\;
\lambda_{\min}(\bm{K}_s)\big(\|\bm{s}\| - n\phi\big),
\]
with $n=3$. Moreover,
\[
    |\bm{s}^\top \bm{\Delta}(t)|
    \le d_{\mathrm{eq}}\,\|\bm{s}\|.
\]

Combining terms,
\[
    \dot{V}
    \le
    -\lambda_{\min}(\bm{K}_s)\big(\|\bm{s}\| - n\phi\big)
    + d_{\mathrm{eq}}\,\|\bm{s}\|.
\]

For sufficiently large $\|\bm{s}\|$, the right-hand side is strictly
negative whenever $\lambda_{\min}(\bm{K}_s) > d_{\mathrm{eq}}$.
Inside the boundary layer $\|\bm{s}\|\le n\phi$, the dynamics remain
bounded, and standard sliding-mode arguments establish that $\bm{s}(t)$
is ultimately bounded in a ball of radius $\delta>0$ around the origin,
with $\delta$ decreasing as $\lambda_{\min}(\bm{K}_s)$ increases and
$\phi$ decreases.
\hfill$\blacksquare$

\subsection{Safety via HOCBF}

Given the HOCBF condition \eqref{eq:hocbf_cond} and the corresponding
linear constraint \eqref{eq:CBF_constraint}, Theorem~1 guarantees
forward invariance of the safe set $\mathcal{C}$ so long as the applied
input $\bm{\tau}$ satisfies \eqref{eq:CBF_constraint} for all $t$.

The projection-based safety filter ensures that if the nominal SMC
control violates this constraint, it is replaced by the closest input
that satisfies it.
Therefore, as long as the constraint set is non-empty at each time, the safe set $\mathcal{C}$ is forward invariant.

\medskip
\noindent
\textbf{Assumption 3 (CBF Feasibility).}
\textit{
For all $t\ge0$, the intersection
$\mathcal{U}\cap\mathcal{C}_u(t)$ is non-empty, where
$\mathcal{U}$ is the actuator box set and
$\mathcal{C}_u(t) = \{\bm{\tau} \mid \bm{A}(t)\bm{\tau}\ge \bm{b}(t)\}$.
}

\medskip
\noindent
Under Assumption~3, the HOCBF conditions can always be enforced.

\subsection{Combined SMC + HOCBF with Projection}

We now consider the full closed-loop system with the projected control
$\bm{\tau} = \bm{\tau}_{\text{safe}}$ obtained from the
projection-based safety filter.

\medskip
\noindent
\textbf{Theorem 3 (Safe Practical Stability).}
\textit{
Assume:
\begin{enumerate}
    \item The conditions of Theorem~2 hold, i.e., the desired trajectories
    $\bm{\eta}_d$, $\bm{\nu}_d$ are bounded and sufficiently smooth and
    the disturbance $\bm{d}(t)$ is matched and satisfies
    $\|\bm{d}(t)\|\le d_{\max}$ for all $t\ge 0$.
    \item The HOCBF condition \eqref{eq:hocbf_cond} is used to construct
    the linear input constraint \eqref{eq:CBF_constraint}.
    \item Assumption~3 (CBF feasibility) holds, i.e., for all $t\ge 0$ the
    intersection $\mathcal{U}\cap\mathcal{C}_u(t)$ is nonempty, where
    $\mathcal{U} = \{\bm{\tau}\mid\bm{\tau}_{\min}\le\bm{\tau}\le\bm{\tau}_{\max}\}$
    and $\mathcal{C}_u(t) = \{\bm{\tau}\mid\bm{A}(t)\bm{\tau}\ge\bm{b}(t)\}$.
\end{enumerate}
Let $\bm{\tau}_{\text{SMC}}$ be the nominal SMC input and
$\bm{\tau}_{\text{safe}}$ the projected input produced by the safety filter,
and define the input deviation
\(
    \bm{\Delta}_u(t) = \bm{\tau}_{\text{safe}}(t)
                       - \bm{\tau}_{\text{SMC}}(t).
\)
Assume further that the feasible set $\mathcal{U}\cap\mathcal{C}_u(t)$ is
compact and that the resulting deviation is uniformly bounded,
$\|\bm{\Delta}_u(t)\|\le \Delta_{\max}$ for all $t\ge 0$.
Then, defining the effective disturbance bound
\[
    d_{\mathrm{eff}}
    := \frac{d_{\max} + \Delta_{\max}}{\lambda_{\min}(\bm{M})},
\]
if the SMC gain matrix $\bm{K}_s$ satisfies
\[
    \lambda_{\min}(\bm{K}_s) > d_{\mathrm{eff}},
\]
the following hold:
\begin{itemize}
    \item[(i)] The safe set 
    $\mathcal{C} = \{\bm{\eta}\mid h(\bm{\eta})\ge 0\}$ is forward invariant.
    \item[(ii)] The sliding surface $\bm{s}$ in \eqref{eq:sdef} is globally
    bounded and ultimately bounded in a neighborhood of the origin, i.e.,
    there exists $\delta>0$ such that
    \(
        \limsup_{t\to\infty} \|\bm{s}(t)\| \le \delta,
    \)
    where $\delta$ can be reduced by increasing 
    $\lambda_{\min}(\bm{K}_s)$ and decreasing $\phi$, subject to actuator
    limits.
\end{itemize}
}

\medskip
\noindent
\textit{Proof:}
(i) \emph{Safety.}
By construction, the projection-based safety filter enforces
$\bm{\tau}_{\text{safe}}(t) \in \mathcal{U}\cap\mathcal{C}_u(t)$ for all $t$,
provided Assumption~3 holds.
In particular, the linear HOCBF constraint \eqref{eq:CBF_constraint}
is satisfied at all times.
Hence the HOCBF condition \eqref{eq:hocbf_cond} holds, and by Theorem~1
the safe set $\mathcal{C} = \{\bm{\eta}\mid h(\bm{\eta})\ge 0\}$ is forward
invariant.

\smallskip
(ii) \emph{Practical stability of $\bm{s}$.}
When the CBF constraint is inactive, the projection leaves the nominal
control unchanged and $\bm{\tau}_{\text{safe}} = \bm{\tau}_{\text{SMC}}$.
In this case, the closed-loop dynamics of $\bm{s}$ reduce to those considered
in Theorem~2, and the sliding variable is ultimately bounded as shown there.

When the CBF constraint is active, the applied input differs from the
nominal one by $\bm{\Delta}_u(t)$, so that
\[
    \bm{\tau}(t) = \bm{\tau}_{\text{safe}}(t)
    = \bm{\tau}_{\text{SMC}}(t) + \bm{\Delta}_u(t).
\]
This additional term enters the $\bm{s}$-dynamics as a matched perturbation
through the plant dynamics:
\[
    \dot{\bm{s}}
    = -\bm{K}_s\,\mathrm{sat}\!\left(\tfrac{\bm{s}}{\phi}\right)
      + \bm{\Delta}_d(t) + \bm{\Delta}_u'(t),
\]
where $\bm{\Delta}_d(t) = \bm{M}^{-1}\bm{d}(t)$ and
$\bm{\Delta}_u'(t) = \bm{M}^{-1}\bm{\Delta}_u(t)$.
By the bounds on $\bm{d}(t)$ and $\bm{\Delta}_u(t)$ and the positive
definiteness of $\bm{M}$, both perturbations are uniformly bounded and
\[
    \|\bm{\Delta}_d(t) + \bm{\Delta}_u'(t)\|
    \le \frac{\|\bm{d}(t)\| + \|\bm{\Delta}_u(t)\|}
               {\lambda_{\min}(\bm{M})}
    \le d_{\mathrm{eff}}
    \quad \forall t\ge 0.
\]
Thus, the total matched disturbance in the $\bm{s}$-dynamics is bounded
by $d_{\mathrm{eff}}$.

Using the same Lyapunov function $V(\bm{s}) = \tfrac{1}{2}\bm{s}^\top \bm{s}$
and the saturation bound $s_i\,\mathrm{sat}(s_i/\phi) \ge |s_i| - \phi$, one
can show, as in Theorem~2, that
\[
    \dot{V}
    \le
    -\lambda_{\min}(\bm{K}_s)\bigl(\|\bm{s}\| - n\phi\bigr)
    + d_{\mathrm{eff}}\,\|\bm{s}\|,
\]
with $n=3$. For $\|\bm{s}\|$ larger than a threshold depending on
$\phi$ and $d_{\mathrm{eff}}$, the right-hand side is strictly negative
whenever $\lambda_{\min}(\bm{K}_s) > d_{\mathrm{eff}}$. Inside the
boundary layer, the dynamics remain bounded, and standard sliding-mode
arguments imply that $\bm{s}(t)$ is ultimately bounded in a ball of
radius $\delta>0$ around the origin, with $\delta$ decreasing as
$\lambda_{\min}(\bm{K}_s)$ increases and $\phi$ decreases.

Together with part (i), this establishes safe practical stability of the
closed-loop system under the combined SMC--HOCBF--projection architecture.
\hfill$\blacksquare$

\subsection{Apparent Adaptiveness of the SMC--CBF Framework}

It is important to emphasize that this mechanism does not constitute true 
adaptation, as no uncertainty model is estimated or updated online, and no 
controller parameters evolve over time. Instead, the SMC--HOCBF structure enables 
a form of \emph{implicit, state-driven adaptiveness}, where safety is enforced 
with varying intensity depending on the proximity to obstacles and the predicted 
evolution of the safety function. This distinguishes the SMC--HOCBF approach from 
learning-based or data-driven adaptive methods, while still providing 
computationally efficient and robust safety guarantees suitable for 
resource-constrained marine platforms.

\section{Simulation Results}

This section presents a comparative study of the navigation with obstacle avoidance between the proposed
SMC--HOCBF controller and a tube-based nonlinear MPC (NMPC) formulation
augmented with control barrier functions, as developed in \cite{TUBEMPC_SYNTAKAS}.
Both controllers are tested on the same 3-DOF marine vessel model,
under identical environmental disturbances (wind, waves, and current),
identical initial and goal states, and identical circular obstacle
configurations.  
The purpose of this comparison is twofold:
(i) to establish a common baseline of robustness through a consistent
tube size, and (ii) to evaluate computation time, safety guarantees, and
trajectory characteristics of both approaches.

\subsection{Matching the Tube Width for Fair Comparison}

The tube-based NMPC controller uses a precomputed constraint tightening
term obtained from extensive Monte Carlo simulations.  
These simulations generate a distribution of reachable deviations between
nominal and disturbed trajectories, from which an empirical tube width is
constructed.  
Because this tube is obtained through repeated stochastic rollouts, it
serves as a \emph{ground truth approximation} of the disturbance-induced
uncertainty envelope around the vessel.

To allow a meaningful comparison, the hyperparameters of the sliding-mode
controller (namely, the matrices $\mathbf{\Lambda}$ and $\mathbf{K}_s$ and the
boundary-layer thickness $\phi$) were tuned so that the resulting
SMC-induced position error bound,
\[
\|e_p(t)\| \le \frac{\phi}{\lambda_{\min}(\mathbf{\Lambda})},
\]
matches the empirical tube radius of the tube-based NMPC.
This ensures that both controllers operate under effectively the same
robustness budget, and that the safe sets constructed using barrier
functions are inflated by comparable amounts.

\subsection{Comparison Under Identical Disturbances}

Both controllers were evaluated under:
\begin{itemize}
    \item time-varying wind forces generated through a spectral model,
    \item second-order wave drift forces with stochastic components,
    \item current forces,
    \item sensor perturbations emulating GPS noise.
\end{itemize}

The SMC--HOCBF controller maintains safety by projecting the nominal SMC
input onto the HOCBF-certified admissible set, while the tube-based NMPC
maintains safety using a tightened feasible region derived from
Monte Carlo rollouts.
Despite their very different design philosophies, both methods achieve
robust collision avoidance and convergence to the goal.

\subsection{Computational Characteristics}

The SMC--HOCBF method operates at significantly lower computational cost:
it requires only algebraic updates and a lightweight projection iteration,
with no need to solve large-scale nonlinear programs.
This makes it suitable for real-time embedded deployment.

However, this reduction in computational burden comes at the cost of
hyperparameter sensitivity.  
The selection of $(\mathbf{\Lambda}, \mathbf{K}_s, \phi)$ strongly affects the
shape of the resulting tube, its symmetry in the $x$ and $y$ directions,
and the aggressiveness of the trajectory near the barrier.
In contrast, the tube-based NMPC relies on a data-driven Monte Carlo
procedure that automatically captures the disturbance distribution and
produces a statistically consistent tube, requiring less manual tuning.

\subsection{Obstacle Avoidance Behavior}

\begin{figure}[t]
    \centering
    \includegraphics[width=0.95\linewidth]{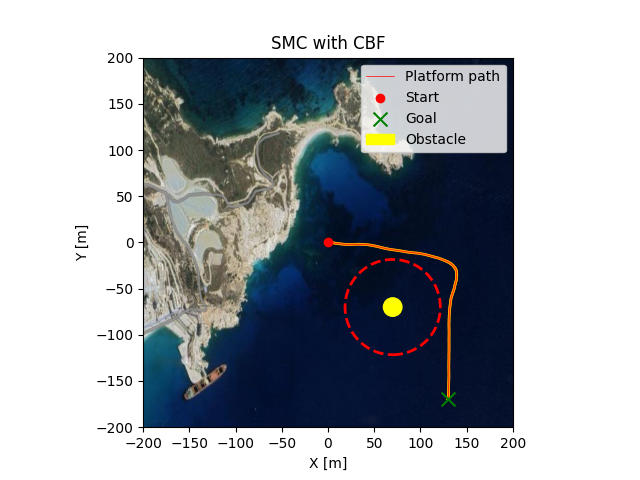}
    \caption{
    Obstacle-avoidance trajectory using the proposed SMC--HOCBF controller.
    }
    \label{fig:smc_obstacle}
\end{figure}

\begin{figure}[t]
    \centering
    \includegraphics[width=0.95\linewidth]{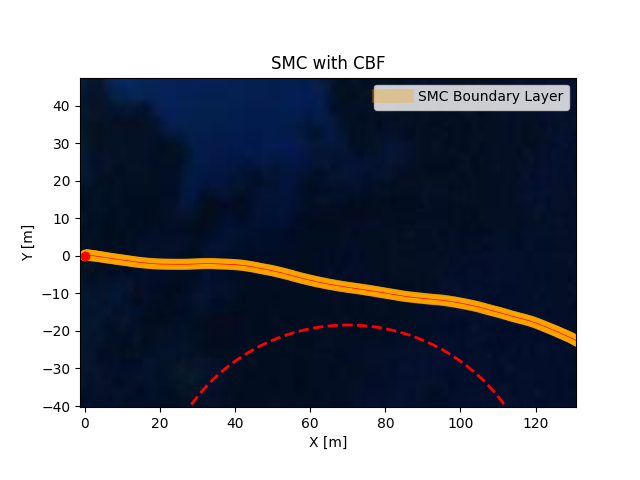}
    \caption{
    SMC Boundary Layer.
    }
    \label{fig:smc_boundary}
\end{figure}

\begin{figure}[t]
    \centering
    \includegraphics[width=0.95\linewidth]{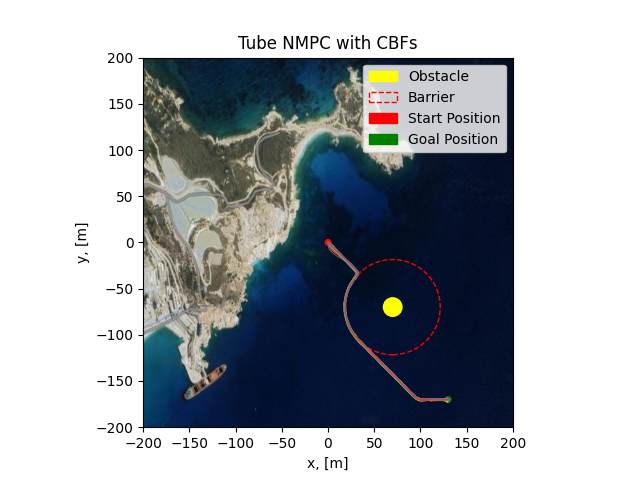}
    \caption{
    Obstacle-avoidance trajectory using the tube-based NMPC with CBF.
    The trajectory exhibits more conservative deviation, reflecting the
    Monte Carlo-derived tube inflation.
    }
    \label{fig:tube_obstacle}
\end{figure}

Fig.~\ref{fig:smc_obstacle} and Fig.~\ref{fig:tube_obstacle} illustrate a
representative obstacle-avoidance scenario for both controllers under
identical disturbances.  
In both cases, the vessel approaches the obstacle boundary and executes a
efficient avoidance maneuver due to the HOCBF-induced safety constraint.
The SMC trajectory tends to ``hug'' the certified boundary more loosely,
due to the  fast corrective dynamics of the sliding mode, whereas the tube-based NMPC yields a more
elegant deviation consistent with its stochastic tube inflation. A depiction of the SMC boundary layer is seen in Fig. \ref{fig:smc_boundary} and as seen the width is equivalent to that of the Tube NMPC.

\begin{figure}[t]
    \centering
    \includegraphics[width=0.95\linewidth]{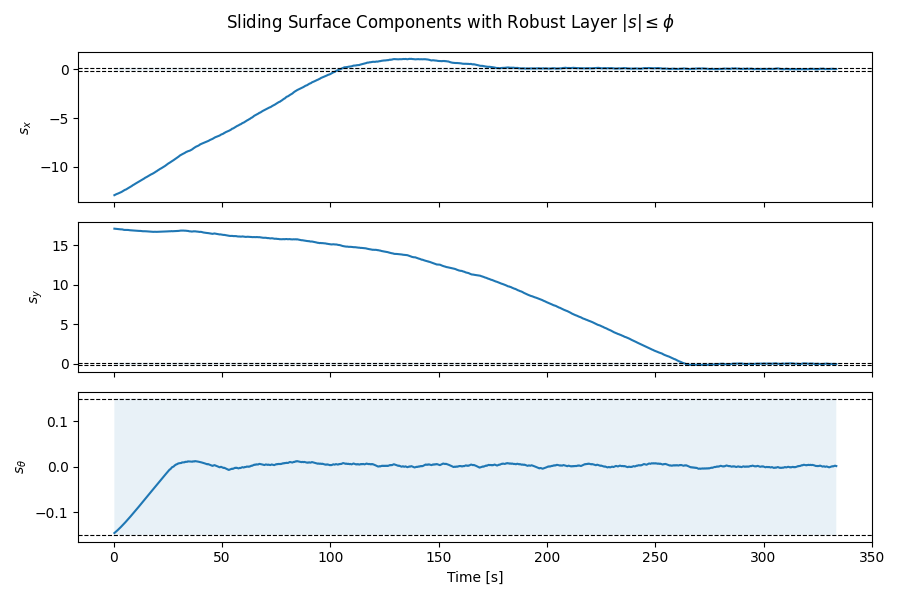}
    \caption{Sliding surface components.}
    \label{fig:smc_components}
\end{figure}

A depiction of the sliding surface components is presented in Fig.~\ref{fig:smc_components}.

Overall, the SMC--HOCBF controller demonstrates comparable safety
performance to the tube-based NMPC while achieving significantly better
computational efficiency.  
By tuning the boundary-layer thickness $\phi$ and the sliding gains,
it can be made to emulate the empirical tube width obtained from the
Monte Carlo-based NMPC, enabling a consistent and fair comparison across
both methodologies.

\section{Conclusion}

This paper introduced a safe control framework combining sliding mode
control, high-order control barrier functions, and a fast projection
operator for 3-DOF marine vessels under disturbances.
The method ensures robust trajectory tracking while maintaining safety
constraints through forward invariance of a CBF-defined safe set.
A Lyapunov-based analysis established practical stability of the sliding
surface and safety of obstacle avoidance.As stated, a notable advantage of the proposed SMC–CBF controller is its very low computational cost compared
to predictive approaches such as tube-based NMPC. Since SMC relies only on algebraic evaluations of the sliding
surface and closed-form expressions of the nominal control law, no nonlinear program (NLP) or numerical optimization
is solved online. The safety filter introduced through the CBF constraint requires only a simple projection onto linear
half-spaces, which is orders of magnitude cheaper than solving a constrained optimal control problem at each sampling
instant. As a result, the SMC–CBF scheme achieves significantly faster update rates, enabling high-frequency control
even under strong disturbances. This stands in contrast to predictive controllers, whose computational effort grows
with the horizon length, number of scenarios, or Monte-Carlo sampling required for robustness.
Thus, for small marine robots and surface vessels with limited onboard computational resources—where execution
speed and computational efficiency are critical—the SMC–CBF framework constitutes a strong candidate for safety-
critical control. Its reliance on closed-form control laws and lightweight constraint enforcement enables robust and safe
operation without the computational burden typically associated with optimization-based methods.
Future work includes extensions to multiple interacting vessels and
experimental validation.

\bibliographystyle{IEEEtran}
\bibliography{bibliography}

\end{document}